\documentstyle[prl,aps,epsf,multicol]{revtex}
\begin{document}
\draft
\title{Effect of zero point phase fluctuations on Josephson tunneling}
\author{Gert-Ludwig Ingold$^{1,2}$ and Hermann Grabert$^3$}
\address{$^1$Institut f\"ur Physik, Universit\"at Augsburg,
Universit{\"a}tsstra{\ss}e~1, D-86135 Augsburg, Germany}
\address{$^2$CEA, Service de Physique de l'Etat Condens\'e, Centre d'Etudes de
Saclay, F-91191 Gif-sur-Yvette, France}
\address{$^3$Fakult\"at f\"ur Physik, Albert-Ludwigs-Universit{\"a}t,
Hermann-Herder-Stra{\ss}e~3, D-79104 Freiburg, Germany}
\date{\today}
\maketitle
\widetext
\begin{abstract}
In the presence of phase fluctuations the dc Josephson effect is
modified and the supercurrent at zero voltage is replaced by a peak at
small but finite voltages. It is shown that at zero temperature this
peak is determined by two complementary expansions of 
finite radius of convergence. The leading order expressions are related 
to results known from the regimes of Coulomb blockade and of macroscopic 
quantum tunneling. The peak positions and the suppression of the
critical current by quantum fluctuations are discussed.
\end{abstract}


\raggedcolumns
\begin{multicols}{2}
\narrowtext
The dc Josephson effect allows a Cooper pair current to flow through a
superconducting tunnel junction in the absence of an external voltage.
The current is determined by the difference $\varphi$ of the condensate
phases on the two sides of the junction through $I=I_c\sin(\varphi)$ and
is limited by the critical current $I_c$. While this feature in the
current-voltage characteristic has zero weight, it acquires a finite width
due to either thermal or quantum fluctuations 
of the phase difference. Here, we analyze the role of
quantum fluctuations and concentrate on the Josephson peak at zero
temperature.

The finite capacitance $C$ of a Josephson junction is a source of
fluctuations of the phase difference $\varphi$ and thus of the broadening 
of the Josephson peak since the charge $Q$ on the capacitance is the 
conjugate variable to $\varphi$ \cite{tinkh96}.
At finite voltage $V$ a current can 
only flow if the tunneling Cooper pairs can loose their excess energy $2eV$. 
This energy can be transferred to the degrees of freedom present in the 
electromagnetic environment of the junction described by the impedance 
of the circuit. In order to observe a peak in the 
current as a function of the voltage, the Josephson junction needs to be
voltage-biased. This has become possible experimentally only very
recently \cite{stein99}.  Since we are interested in the behavior close to the
ideal Josephson peak, typical voltages $V$ are much smaller than the gap
voltage $\Delta/2e$ and quasiparticle excitations can be neglected. 

\begin{figure}
\begin{center}
\leavevmode
\epsfxsize=0.2\textwidth
\epsfbox{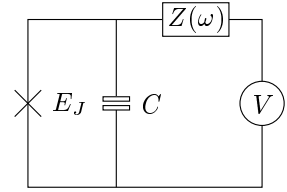}
\end{center}
\caption{Josephson junction characterized by the Josephson energy $E_J$
and capacitance $C$ coupled to an ideal voltage source $V$ via an
external impedance.}
\label{fig:circuit}
\end{figure}

A minimal circuit displaying quantum fluctuations of the phase is
shown in fig.~\ref{fig:circuit} and may be described in terms of the
Hamiltonian \cite{ingol91}
\begin{eqnarray}
\label{eq:hamiltonian}
H &=& \frac{Q^2}{2C} - E_J\cos(\varphi)\\
&&+ \sum_{n=1}^{\infty}\left[
\frac{q_n^2}{2C_n}+\left(\frac{\hbar}{2e}\right)^2\frac{1}{2L_n}
\left(\frac{2e}{\hbar}Vt-\varphi-\varphi_n\right)^2\right].
\nonumber
\end{eqnarray}
The first term corresponds to the charging energy and introduces an
energy scale $E_c=2e^2/C$. The second term 
describes the tunneling of Cooper pairs through\break
\noindent the junction. The
Josephson coupling energy $E_J$ is related to the critical current by
$I_c=2eE_J/\hbar$. The third term describes the coupling of the
junction to an external impedance modeled by a set of $LC$-circuits and
also takes into account an applied voltage $V$. 

Summing the perturbative expansion in the Josephson coupling
to all orders, one finds for the equilibrium Cooper pair 
current \cite{grabe98}
\begin{eqnarray}
I &=& 2e\sum_{n=1}^{\infty}\left(i\frac{E_J}{2\hbar}\right)^{2n}
\sum_{\{\zeta, \eta\}}\left(\prod_{k=1}^{2n-1}\eta_k\right)\zeta_0\nonumber\\
&&\times\int_0^{\infty}dt_1\dots\int_0^{t_{2n-2}}dt_{2n-1} \exp(\Gamma)
\label{eq:pertseries}
\end{eqnarray}
where the exponent in the integrand is given by
\begin{equation}
\Gamma=i\frac{2e}{\hbar}V\sum_{k=0}^{2n-1}\zeta_kt_k-
\sum_{k=1}^{2n-1}\sum_{l=0}^{k-1}\zeta_k\zeta_lJ[\eta_k(t_k-t_l)].
\label{eq:exponent}
\end{equation}
The sums over $\zeta_k$ and $\eta_k$ run over the values $\pm1$ with
the constraint $\sum_{k=0}^{2n-1}\zeta_k=0$. 

This result depends on the phase autocorrelation function in the absence
of tunneling \cite{ingol91} which at zero temperature is given by
\begin{equation}
J(t) = 2\int_0^{\infty}\frac{d\omega}{\omega}\frac{{\rm Re}
Z_t(\omega)}{R_Q}(e^{-i\omega t}-1).
\label{eq:jvont}
\end{equation}
It is entirely determined by the total impedance $Z_t(\omega) =
[i\omega C+1/Z(\omega)]^{-1}$ seen by 
the junction. In the following, we assume a purely ohmic external resistance 
$R$ thereby neglecting features in the impedance,
like resonances in a transmission line,  which might lead to 
additional structure in the $I$--$V$--curve 
\cite{ingol94,holst94}. One then has
\begin{equation}
\frac{{\rm Re}Z_t(\omega)}{R_Q} = \frac{\rho}{1+(\omega/\omega_R)^2}
\label{eq:totimp}
\end{equation}
where $\rho=R/R_Q$ with the resistance quantum $R_Q=h/4e^2$. The total
impedance is cut off at a frequency $\omega_R=1/RC$ due to the junction
capacitance. For sufficiently long times the correlation function is
given by
\begin{equation}
J(t) = -2\rho\left[\ln\left(\omega_R\vert t\vert\right) +\gamma
+i\frac{\pi}{2}{\rm sign}(t)\right]
\label{eq:jvont0}
\end{equation}
where $\gamma=0.5772\dots$ is the Euler constant. 

Introducing a dimensionless time $2eVt/\hbar$ and making use of the
constraint on the $\zeta_k$, one finds that for the
correlation function (\ref{eq:jvont0}) each term of the perturbation
series (\ref{eq:pertseries}) depends on the Josephson coupling only through
the combination $(E_J/V^{1-\rho})^{2n}$. Therefore, the perturbative
expansion in $E_J$ will finally lead to a power series in the applied
voltage $V$. This is a consequence of the special form of the
correlation function (\ref{eq:jvont}) at zero temperature.

It is instructive to first discuss the problem in several limits where results
are already available. We start with the regime of classical phase
diffusion which corresponds to taking the limit $\rho = 0$. Writing the
expansion (\ref{eq:pertseries}) in terms of a continued fraction 
\cite{grabe98} one obtains
\begin{equation}
\frac{I}{I_c} = \frac{V}{RI_c}-
\left[\left(\frac{V}{RI_c}\right)^2-1\right]^{1/2}
\Theta\left(\frac{V}{RI_c}\right)
\label{eq:ivanzilb}
\end{equation}
where $\Theta(x)$ is the Heaviside step function. This corresponds to
the zero temperature limit of the result obtained by Ivanchenko and
Zil'berman \cite{ivanc68}. The current-voltage characteristic 
(\ref{eq:ivanzilb}) starts with an ohmic line and displays a
cusp at $V=RI_c$. A further increase of the voltage results in a
decreasing current.

As discussed above, an expansion in the Josephson coupling energy at 
$\rho=0$ amounts to an expansion in $1/V$. The series (\ref{eq:pertseries}) 
corresponds to the Taylor expansion 
of (\ref{eq:ivanzilb}) given by
\begin{equation}
I = \frac{1}{2\pi^{1/2}}\frac{V}{R}\sum_{n=1}^{\infty}
\frac{\Gamma(n-1/2)}{\Gamma(n+1)}\left(\frac{V}{RI_c}\right)^{-2n}.
\label{eq:ivanzilbexp}
\end{equation}
This series has a finite radius of convergence and is restricted to 
$\vert V\vert > RI_c$. The limit of convergence just coincides 
with the position of the cusp. Hence, we see that for small $\rho$ and low
temperatures the expansion (\ref{eq:pertseries}) will only
converge for sufficiently large voltages.

To examine this further, we now allow for finite $\rho$ 
and turn to the regime of charging effects 
where $E_c\gg E_J$. Within the standard theory of environmental
effects on Coulomb Blockade (CB) \cite{ingol91}, 
tunneling is treated perturbatively and the current-voltage 
characteristics is obtained from the leading term ($n=1$) of 
the series (\ref{eq:pertseries}) 
\cite{averi90}. For an environment with an ohmic low frequency component, one 
finds a zero bias anomaly of the Cooper pair current \cite{devor90}
\begin{equation}
I = \frac{\pi^{1/2}\rho}{2\Gamma(\rho)\Gamma(\rho+1/2)}\frac{(\pi
E_J)^2}{(\hbar\omega_R e^{\gamma})^{2\rho}}\frac{V}{R}(eV)^{2\rho-2}.
\label{eq:ivpe}
\end{equation}
While for $\rho > 1$ this result describes the suppression of the current
by the Coulomb blockade effect, for $\rho < 1$ it corresponds 
to a divergent zero bias conductance. 
One might hope that higher order terms regularize 
the divergence, but the discussion of the classical phase 
diffusion limit suggests that for $\rho<1$ the series is divergent for small
$V$. We will show below that this is indeed the case.

Now, for small $\rho$ and small voltages, corrections to the linear 
part of the current-voltage characteristic
(\ref{eq:ivanzilb}) arise from Macroscopic
Quantum Tunneling (MQT). In this regime the Josephson junction is mostly in
its zero voltage state and the voltage drop occurs at the resistor.
Occasionally, a phase slip will cause a finite voltage across the
junction leading on average to a finite dc contribution. 
In the overdamped limit
$2\pi^2\rho^2 E_J \ll E_c$ the current-voltage characteristic is given
by \cite{korsh87}
\begin{equation}
I = \frac{V}{R}\left(1 - \frac{\pi}{2 e^{1/2}
(\pi\rho)^{5/2+2/\rho}} \frac{E_c^2}{E_J^{2/\rho}}
(eV)^{2/\rho-2}\right).
\label{eq:korshunov}
\end{equation}
Note that the zero bias differential conductance goes to zero
for $\rho<1$ where we found a divergence within CB theory. On
the other hand, the zero bias differential conductance of
(\ref{eq:korshunov}) diverges for $\rho>1$.

In order to reconcile these findings we make use of the analogy between
a Josephson junction and a damped particle in a
periodic potential. Schmid \cite{schmi83} has noted that the
regions of small and large $\rho$ are related by a self--duality of the
model. Further progress 
\cite{fendl95,weiss96} in the calculation of the
mobility of the damped particle has been based on the thermodynamic 
Bethe ansatz \cite{zamol90}. Duality has also been exploited in the context 
of the fractional quantum Hall effect \cite{fendl95,chamo97}.

It should be emphasized that self--duality relies on
strictly ohmic damping. In contrast, the spectrum (\ref{eq:totimp}) 
has a cutoff frequency $\omega_R$. This does not spoil duality
in the long-time limit of the correlation
function (\ref{eq:jvont0}) where the dependence on $\omega_R$ can
be absorbed in an effective voltage scale introduced below. 
The long-time limit
restricts us to small voltages with $eV\ll\hbar\omega_R$ or equivalently 
$V/RI_c\ll E_c/\pi^2\rho^2E_J$. For the typical case of small environmental 
impedances, voltages of interest are of order $RI_c$, and 
then a rather wide range of ratios $E_c/E_J$ is allowed. Below we will
show how the $I$--$V$--curve can be extended beyond the
voltage limit imposed by the strictly ohmic approximation.

As a consequence of duality, the current voltage characteristics
can be obtained from an integral representation \cite{weiss96,fendl98}
which implies two complementary expansions describing the
zero temperature behavior. In the scaling limit the CB
series (\ref{eq:pertseries}) takes the form
\begin{equation}
I = \frac{V}{R}\sum_{n=1}^{\infty}c_n(\rho)
\left(\frac{V}{V_0}\right)^{2(\rho-1)n}.
\label{eq:exi}
\end{equation}
On the other hand, the MQT
series, with the leading order terms (\ref{eq:korshunov}), reads
\begin{equation}
I = \frac{V}{R}\left(1-\sum_{n=1}^{\infty}c_n(1/\rho)
\left(\frac{V}{V_0}\right)^{2(1/\rho-1)n} \right).      
\label{eq:exii}
\end{equation}
The coefficients of these expansions are determined by duality and given by
\begin{equation}
c_n(\rho) = (-1)^{n-1}\frac{\Gamma(1+\rho n)\Gamma(3/2)}{\Gamma(1+n)
\Gamma(3/2+(\rho-1)n)}
\label{eq:cn}
\end{equation}
and the voltage scale is set by
\begin{equation}
\label{eq:v0}
V_0 = \frac{\pi E_J}{e}\left[\Gamma(\rho)\left(\frac{e^{\gamma}}{\pi^2\rho} 
\frac{E_c}{E_J}\right)^{\rho}\right]^{1/(\rho-1)}.
\label{eq:vnull}
\end{equation}

\begin{figure}
\begin{center}
\leavevmode
\epsfxsize=0.42\textwidth
\epsfbox{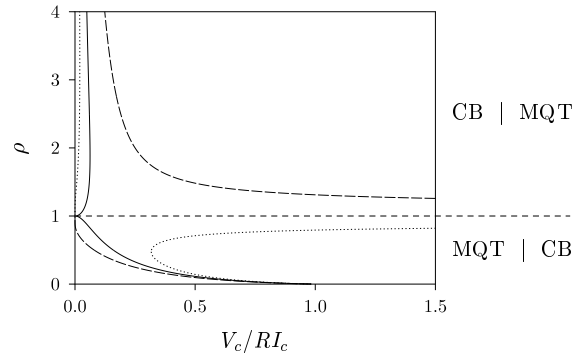}
\end{center}
\caption{Range of convergence of the two expansions
(\protect\ref{eq:exi}) and (\protect\ref{eq:exii}) for
$(e^{\gamma}/\pi^2)(E_c/E_J) = $ 0.5, 1 and 2 (dotted, full and dashed
line, respectively). Below the horizontal dashed line $\rho=1$, the series 
(\protect\ref{eq:exi}) converges to the right of the curves while above 
$\rho<1$ it converges to their left. The opposite holds for the series 
(\protect\ref{eq:exii}).}
\label{fig:convergence}
\end{figure}

The two expansions (\ref{eq:exi}) and (\ref{eq:exii}) have a finite
radius of convergence which can be expressed in terms of a critical
voltage
\begin{equation}
V_c =  V_0\left(\vert 1-\rho\vert\rho^{\rho/(1-\rho)}\right)^{1/2}.
\label{eq:vcrit}
\end{equation}
For $\rho<1$ the series (\ref{eq:exi}) converges for $V>V_c$ and thus is 
a large voltage expansion, as discussed above, 
while the series (\ref{eq:exii}) converges for
low voltages $V<V_c$. The role of the expansions is interchanged above
$\rho=1$ where (\ref{eq:exi}) yields a low voltage expansion while
(\ref{eq:exii}) determines the large voltage behavior. The situation is
illustrated in fig.~\ref{fig:convergence} where on the right side
it is indicated which expansion converges on which side of the curves.

\begin{figure}
\begin{center}
\leavevmode
\epsfxsize=0.42\textwidth
\epsfbox{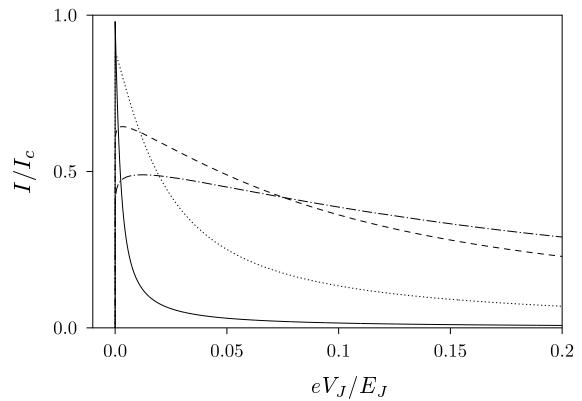}
\end{center}
\caption{Zero temperature Cooper pair current-voltage characteristics
for $(e^{\gamma}/\pi^2)(E_c/E_J)=1$ and $\rho=$ 0.001, 0.01, 0.05 and
0.1 shown as full, dotted, dashed, and dashed-dotted line, respectively.}
\label{fig:ivs}
\end{figure}

As we have already seen for the case $\rho=0$, the current-voltage
characteristic displays a peak. This is still true for finite $\rho$
as seen from fig.~\ref{fig:ivs} where we present $I$--$V$--curves for 
various values of $\rho$. Here, we show the current as a function
of the voltage $V_J$ across 
the Josephson junction which is related to the externally applied voltage by
$V_J=V-RI$. As can be seen from the figure, for finite external
impedance, the $I$--$V$--curve corresponds to a peak of
finite width and a maximum current suppressed with respect to the
critical current in the absence of fluctuations. As $\rho$ is decreased, the 
peak narrows and the usual Cooper pair current at
zero voltage builds up.

\begin{figure}
\begin{center}
\leavevmode
\epsfxsize=0.42\textwidth
\epsfbox{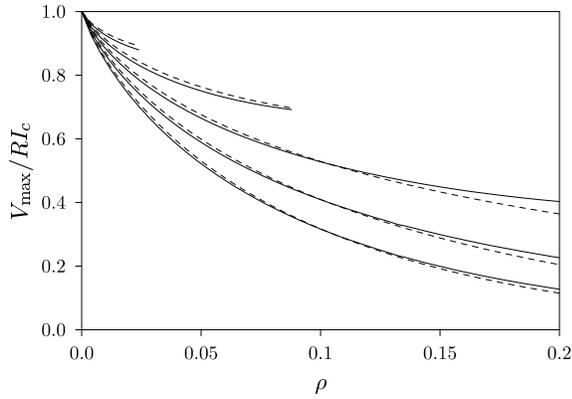}
\end{center}
\caption{Comparison of peak positions (full lines) and the critical
voltage (\ref{eq:vcrit}) for $(e^{\gamma}/\pi^2)(E_c/E_J)=$ 0.01, 0.1,
1, 10, and 100 increasing from the upper to the lower curves. 
Results are only shown for values of $\rho$ which satisfy $eV_{\rm max}\le
10\hbar\omega_R$.}
\label{fig:vmax}
\end{figure}

Due to the complicated form of the expansions (\ref{eq:exi}) and
(\ref{eq:exii}) an analytic determination of the position and height 
of the peak in the current-voltage characteristics is in general not
possible. However, it turns out that to\break
\noindent a very good approximation the 
peak position is given by the critical voltage (\ref{eq:vcrit}). 
The quality of this approximation can be seen from
fig.~\ref{fig:vmax} where we compare it (dashed lines) with the peak
positions determined numerically (full lines).  The lines have been
restricted to values of $\rho$ for which $eV_{\rm max}\le
10\hbar\omega_R$ in order to ensure the applicability of the theory.
Likewise, the maximum current can quite reliably be estimated by 
$I_{\rm max} \sim V_c/R$. For more precise
results, a numerical evaluation of (\ref{eq:exi}) and (\ref{eq:exii}) is
required which does not present special problems.

As already emphasized, the range of validity of the expansions 
(\ref{eq:exi}) and
(\ref{eq:exii}) is restricted by the assumption of strictly ohmic
damping. However, the current-voltage characteristics can be extended
to larger voltages if CB theory
yields a good description at the limit of validity. 
One may then continue the $I$--$V$--curve to larger
voltages by using the result of CB theory with the full 
frequency--dependent impedance (\ref{eq:totimp}). 

To illustrate this point we consider the case
$\rho=1/2$ which for strictly Ohmic damping allows for an exact 
solution \cite{guine85,weiss88}. Summing up the two
expansions (\ref{eq:exi}) and (\ref{eq:exii}), one finds in both cases
\begin{equation}
\frac{I}{I_c} = \frac{\pi E_J}{2E_c e^{\gamma}}\text{arctan}
\left(\frac{2 E_c e^{\gamma}}{\pi E_J}\frac{V}{RI_c}\right).
\label{eq:ivhalf}
\end{equation}
Obviously, this result no longer describes a
peak structure as it was the case for $\rho < 1/2$. 
This is however an artifact of the  assumption of ohmic damping. 
For $\rho = 1/2$ the plateau value of (\ref{eq:ivhalf}) just
corresponds to the value (\ref{eq:ivpe}) given by CB theory. 
Beyond the validity of the expansions (\ref{eq:exi}) and
(\ref{eq:exii}) one may thus use the cutoff--dependent leading 
order term of the series (\ref{eq:pertseries}) which describes 
the decrease of the current
for larger voltages and leads again to a peak in the $I$--$V$--curve.
For $E_c\gg E_J$, (\ref{eq:ivhalf}) reaches its plateau value for very
small voltages. Then, CB theory only fails for
voltages below $(E_J/E_c) RI_c$. Quite generally, 
for $\rho <1$ and $E_c\gg E_J$ the result of CB theory with the 
exact impedance can be employed except for small voltages. There,
however, the strictly ohmic approximation is
appropriate and the results discussed above can be used.

For $\rho>1$, CB theory describes the low voltage behavior. For example
for $\rho=2$, the case dual to $\rho=1/2$, one finds 
\begin{equation}
\frac{I}{I_c} = \frac{V}{RI_c}-\left(\frac{E_c e^{\gamma}}{4\pi^2
E_J}\right)^2 \text{arctan}\left[\left(\frac{4\pi^2 E_J}{E_c e^{\gamma}}
\right)^2\frac{V}{RI_c}\right]
\label{eq:ivtwo}
\end{equation}
which agrees with (\ref{eq:ivpe}) to leading order in $V$. This result
for strictly Ohmic damping
diverges for large $V$ but is in fact regularized  by the cutoff in the
impedance. Thus for $\rho>1$ and $E_c\gg E_J$ the entire peak 
in the $I$--$V$--curve is determined by CB theory.

In conclusion, we have determined the shape of the Josephson current
peak in the presence of quantum fluctuation. We have shown that the
low voltage behavior can be determined within the approximation of
a self--dual model with strictly Ohmic impedance. The results of the dual model
have been connected with the phenomena of Coulomb blockade and macroscopic 
quantum tunneling. This allows for the calculation of the $I$-$V$-curve
for experimentally relevant frequency-dependent impedances in a large
range of parameters accessible by state--of--the--art technology. 

The authors would like to thank M.\ H.\ Devoret, D.\ Esteve, and A.\
Steinbach for many inspiring discussions. One of us (GLI) is grateful to
the SPEC for hospitality and the Volkswagen-Stiftung for financial
support during a stay at the CEA Saclay. The other author (HG) was
supported by the DAAD through PROCOPE.

\end{multicols}

\begin{references}

\bibitem{tinkh96}
M.~Tinkham, {\it Introduction to Superconductivity} (Mc\-Graw-Hill, New
York, 1996).

\bibitem{stein99} A.\ Steinbach, D.\ Esteve, M.\ H.\ Devoret, private
communication.

\bibitem{ingol91}
G.-L.\ Ingold and Yu.\ V.\ Nazarov, in : {\sl Single Charge Tunneling},
ed.\ by H.\ Grabert and M.\ H.\ Devoret, NATO ASI Ser.\ B, Vol.\ {\bf 294}
(Plenum, 1991).

\bibitem{grabe98}
H.\ Grabert, G.-L.\ Ingold, and B.\ Paul, Europhys.\ Lett.\ {\bf 44}, 360
(1998).

\bibitem{ingol94}
G.-L.\ Ingold, H.\ Grabert and U.\ Eberhardt, Phys.\ Rev.\ B {\bf 50},
395 (1994).

\bibitem{holst94}
T.\ Holst, D.\ Esteve, C.\ Urbina, and M.\ H.\ Devoret, Physica B
{\bf 203}, 397 (1994);
T.\ Holst, D.\ Esteve, C.\ Urbina, and M.\ H.\ Devoret, Phys.\ Rev.\
Lett.\ {\bf 73}, 3455 (1994).

\bibitem{ivanc68}
Yu.\ M.\ Ivanchenko and L.\ A.\ Zil'berman, Zh.\ Eksp.\ Teor.\ Fiz.\ {\bf
55}, 2395 (1968) [Sov.\ Phys.\ JETP {\bf 28}, 1272 (1969)].

\bibitem{averi90}
D.\ V.\ Averin, Yu.\ V.\ Nazarov, and A.\ A.\ Odintsov, Physica B {\bf
165\&166}, 945 (1990).

\bibitem{devor90}
M.\ H.\ Devoret, D.\ Esteve, H.\ Grabert, G.-L.\ Ingold, H.\ Pothier,
and C.\ Urbina, Phys.\ Rev.\ Lett.\ {\bf 64}, 1824 (1990).

\bibitem{korsh87} 
S.\ E.\ Korshunov, Zh.\ Eksp.\ Teor.\ Fiz.\ {\bf 92}, 1828 (1987)
[Sov.\ Phys.\ JETP {\bf 65}. 1025 (1987)].

\bibitem{schmi83}
A.\ Schmid, Phys.\ Rev.\ Lett.\ {\bf 51}, 1506 (1983).

\bibitem{fendl95}
P.\ Fendley, A.\ W.\ W.\ Ludwig, and H.\ Saleur, Phys.\ Rev.\ Lett.\ {\bf
75}, 2196 (1995);
P.\ Fendley, A.\ W.\ W.\ Ludwig, and H.\ Saleur, Phys.\ Rev.\ B {\bf 52},
8934 (1995).

\bibitem{weiss96}
U.\ Weiss, Solid State Commun.\ {\bf 100}, 281 (1996);
U.\ Weiss, in: {\sl Tunneling and its Implication}, ed.\ by H.\ Cerdeira,
A.\ Ranfagni, and L.\ S.\ Schulman (World Scientific, 1997).

\bibitem{zamol90}
Al.\ B.\ Zamolodchikov, Nucl.\ Phys.\ B {\bf 342}, 695 (1990).

\bibitem{chamo97}
C.\ de C.\ Chamon and E.\ Fradkin, Phys.\ Rev.\ B {\bf 56}, 2012 (1997).

\bibitem{fendl98}
P.\ Fendley, Adv.\ Theor.\ Math.\ Phys.\ {\bf 2}, 987 (1998).

\bibitem{guine85}
F.\ Guinea, Phys.\ Rev.\ B {\bf 32}, 7518 (1985).

\bibitem{weiss88}
U.\ Weiss, M.\ Wollensak, Phys.\ Rev.\ B {\bf 37}, 2729 (1988).
\end{references}
\end{document}